# Astigmatic Speckle-learned OAM Shift Keying and OAM Multiplexing

**Trishita Das[1], Manas Ranjan Pandit[1], Venugopal Raskatla[1,2], Purnesh Singh Badavath[1], Vijay Kumar[1]***

1 National Institute of Technology Warangal, Department of Physics, Hanamkonda, Telangana, 506004, India

2 Optoelectronics Research Centre and Centre for Photonic Metamaterials, University of Southampton, Highfield, Southampton, SO17 1BJ, United Kingdom

*vijay@nitw.ac.in

**Abstract**

Orbital angular momentum (OAM)-carrying beams have gained significant attention in recent years due to their unique properties and potential to improve spectral efficiency and data transmission rates in optical communication systems. However, fully exploiting the capabilities of the entire OAM mode spectrum remains challenging. The emergence of AI-driven OAM mode identification has revolutionized the demultiplexing process within optical communication channels. OAM beams with different orders are orthogonal, allowing each beam to serve as a distinct signal carrier. Combining multiple OAM beams can effectively enhance channel capacity. In this paper, we adopt speckle-learned demultiplexing to demultiplex OAM beams via its speckle pattern that is more resilient to alignment and noise. However, the use of only non-intensity degenerate beams limits the utilization of multiplexing resources. This approach aims to fully leverage the full spectrum of OAM beams by introducing astigmatism in far-field speckle patterns using a tilted spherical convex lens. We then conduct a comprehensive analysis of two innovative information encoding techniques: OAM shift keying and OAM multiplexing. We successfully demonstrate an optical communication link encoded using both OAM shift keying and OAM multiplexing, followed by accurate decoding via speckle-learned demultiplexing.

## 1. Introduction

Structured light beams, imbued with orbital angular momentum (OAM), have piqued the interest of the optical communication community due to their unique properties and capabilities[1]. In 1992, Allen's pioneering work[2] demonstrated that structured light beams characterized by helical phase components (expressed as $e^{il\varphi}$, where '$l$' denotes the topological charge and $\varphi$ signifies the azimuthal angle) carry OAM with a magnitude of $l\hbar$ per photon, directed along the propagation axis. The topological charge '$l$' denotes the number of phase shifts occurring across the beam, effectively dictating the OAM values inherent in each photon. Notably, OAM beams of distinct modes exhibit orthogonality, rendering them capable of functioning as multiple independent information channels in optical communication scenarios[3-5].

However, in the realm of Free-Space Optical (FSO) communication, it is essential to recognize that higher-order OAM modes exhibit greater divergence compared to their lower-order counterparts. This divergence poses challenges when attempting to capture the entire mode using a receiver aperture of limited size, resulting in increased power loss. Diverse techniques are currently under exploration to address this constraint, focusing on the identification of intensity degenerate or mutual conjugate modes to fully harness the OAM spectrum. Traditionally, the determination of such modes has been based on the orientation characteristics of the patterns generated through interference[6-7] and diffraction methodologies while recent approaches employ machine learning. The traditional machine learning methods utilizing the[8-10] direct mode intensity images[11] are very sensitive to alignment and need to capture the whole mode for their classification making it challenging to identify the original modes and decode the encoded information accurately. To overcome this problem, we have used Speckle-Learned CNN [12-16] for the recognition of the intensity degenerate OAM beams through their corresponding astigmatic far-field speckle patterns incorporated by a tilted spherical convex lens[17-18]. Hence enhanced robustness in the face of noise and alignment challenges by employing the entire spectrum of OAM modes.

In the later part of the research paper, we delve into a comprehensive analysis of two innovative approaches for information encoding: OAM Shift Keying and OAM Multiplexing. Shift Keying dynamically modifies OAM states for efficient data transmission while Multiplexing[19] leverages lower-order OAM superposition (OAM-SP) modes to avoid cross-talk among higher-order OAM modes. Finally, to demonstrate the proof of concept for its practical application, we have simulated an optical communication link for both the above-mentioned approaches where we have encoded alphabets and numbers using OAM shift keying as well as using OAM multiplexing, and faithfully reconstructed with Speckle-Learned CNN emphasizing the efficiency and reliability of Speckle-Learned CNN in data demultiplexing. The present approach substantially boosts channel capacity, increases information transfer rates, and enhances the reliability of free-space optical communication.

## 2. Optical Angular Momentum Modes

2.1 *Laguerre–Gaussian ($LG_{p,l}$) Modes*

The optical modes having OAM are the solutions of the free space paraxial wave equation. LG beams, which possess optical vortex properties, are solutions to the paraxial Helmholtz equation in the cylindrical coordinates system. The mathematical expression of the $LG_{p,l}$ beam with radial mode index p and azimuthal mode index l is given by,

$$U_{p,l}(\rho, \varphi, z) = A_0 \left[\frac{w_0}{w(z)}\right] \left(\frac{\rho}{w(z)}\right)^{|l|} L_p^{|l|}\left(\frac{2\rho^2}{w^2(z)}\right) \times \exp\exp\left(\frac{-\rho^2}{w^2(z)}\right) \exp\left[-i\left(kz + k\frac{\rho^2}{2R(z)}\right)\right]$$

$$\times \exp\exp(il\varphi) \times \exp[i(|l| + 2p + 1)\varphi(z)] \quad (1)$$

where $A_0$ is the amplitude, $R(z)$ is the wavefront curvature of the beam, $w(z)$ is the effective width of the beam, $w_0$ is the beam waist at $z = 0$, $\varphi(z)$ is the Gouy phase shift, $p$ is a nonnegative integer, $l$ is an integer and $L_p^{|l|}$ is an associated Laguerre function of order $p$ and $l$. For a given value of $p$, Eq. (1) gives the identical intensity profiles for modes $LG_{p,l}$ and $LG_{p,-l}$ since they are mutual conjugate modes. Each LG beam serves as a distinct channel carrier due to the orthogonality between OAM beams of a different order.

2.2 *Intensity Degenerate or Mutual Conjugate Modes*

The superposition of any $LG_{p,l}$ modes with p = 0 are expressed as,

$$SM_{\{l\}}(r) = \sum_l c_l \, LG_{0,l}(r) \quad (2)$$

Where subscript $\{l\}$ denotes all the contributing $LG_{0,l}$ modes and $c_l$ is their corresponding weight. The modes $SM_{\{l\}}$ and $SM_{\{-l\}}$ are *intensity degenerate* or *mutual conjugate modes*.

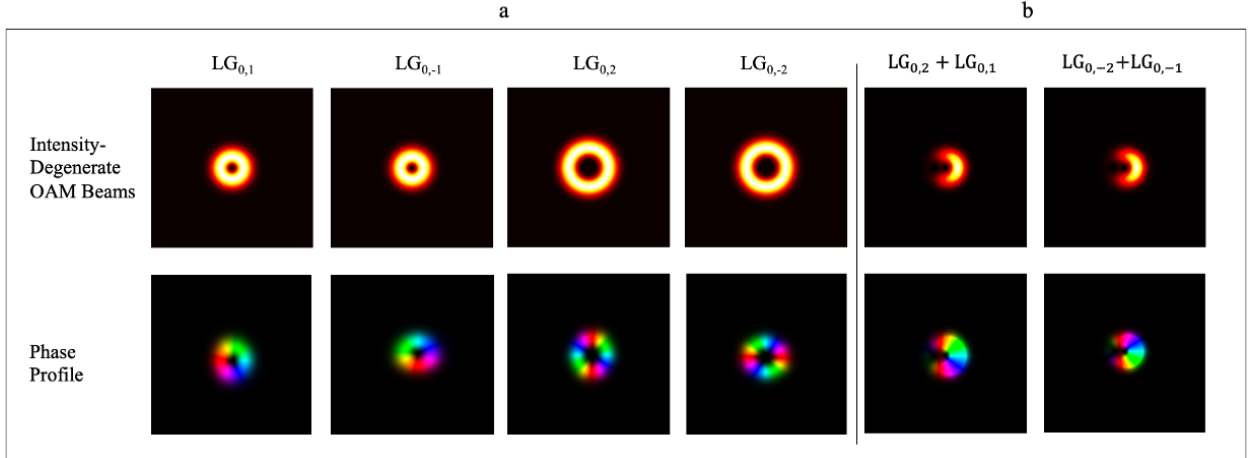

Figure A. Intensity and phase profile of the intensity degenerate modes and their respective far-field intensity speckle images of (a) $LG_{p,l}$ and (b) $SM_{\{l\}}$

In the upcoming section, we will provide a detailed overview of our implementation of OAM shift keying and OAM multiplexing. Subsequently, we will conduct a comprehensive comparison between these two methods, offering insights into their respective advantages and performance characteristics.

## 1.2 OAM Shift Keying

The *OAM shift-keying* (SK) is a method of information encoding that involves the rapid switching of OAM modes. The OAM states are changed over time according to the data strings, essentially mapping data bits to specific OAM states. Each OAM state, in simple terms, represents a single piece of information, like a 0 or a 1. What's fascinating is that in theory, we can have an unlimited number of these OAM states, which means we can transmit a vast amount of data using this method.

$$\text{For transmitting n-bits of information , No. of modes required} = n$$

So in order to increase the number of channels or coding modes, higher-order OAM beams are required. One of the main challenges of using higher-order modes is they are more susceptible to perturbations and atmospheric turbulence. That can result in power loss and mode crosstalk, which can be detrimental to the overall quality of the communication link. Now to avoid crosstalk among higher-order modes we have employed a scheme that will utilize the full spectrum of OAM beams. We had considered $2^5 = 32$ $LG_{p,l}$ modes (including intensity degenerate modes) with $p = 0$ and $l = 1$ to $\pm 16$ to encode the 5-bit information using OAM-SK. Figure A. (I) shows the 5-bit information encoding scheme using OAM-SK.

## 1.3 OAM Multiplexing

The second approach is termed as *OAM multiplexing*, where each OAM state serves as an independent data carrier or channel. Rather than changing the OAM states over time as in OAM-SK, OAM multiplexing leverages the orthogonality of these states. Multiple channels of information are transmitted simultaneously using different OAM states. To ensure that these channels are distinguishable from one another, various properties of light, such as amplitude and phase, are modulated along with the OAM states. This approach significantly increases the total data capacity of the communication system by utilizing the unique OAM superposition.

To realize a 5-bit multiplexing scheme for information encoding, we considered only six $LG_{p,l}$ beams ($p = 0, l = 0 \pm 1, \pm 2, \& 3$) and constructed thirty-two OAM beams superposed. Figure A. (II) shows the 5-bit information encoding scheme using OAM demultiplexing.

For transmitting n-bits of information, No. of modes required = $\sqrt{n}$.

Thus OAM space division multiplexing has a higher spectral efficiency than OAM Shift Keying.

| I. **OAM Shift-Keying** | | II. **OAM Multiplexing** | | | | | | |
|---|---|---|---|---|---|---|---|---|
| Notations | Intensity Degenerate LG Modes | $LG_{0,3}$ | $LG_{0,-2}$ | $LG_{0,2}$ | $LG_{0,-1}$ | $LG_{0,1}$ | Superposed LG Modes | Notations |
| A | $LG_{0,-16}$ | 0 | 0 | 0 | 0 | 0 | Gaussian Beam | A |
| B | $LG_{0,-15}$ | 0 | 0 | 0 | 0 | 1 | $LG_{0,1}$ | B |
| C | $LG_{0,-14}$ | 0 | 0 | 0 | 1 | 0 | $LG_{0,-1}$ | C |
| D | $LG_{0,-13}$ | 0 | 0 | 0 | 1 | 1 | $LG_{0,1} + LG_{0,-1}$ | D |
| E | $LG_{0,-12}$ | 0 | 0 | 1 | 0 | 0 | $LG_{0,2}$ | E |
| F | $LG_{0,-11}$ | 0 | 0 | 1 | 0 | 1 | $LG_{0,2} + LG_{0,1}$ | F |
| G | $LG_{0,-10}$ | 0 | 0 | 1 | 1 | 0 | $LG_{0,2} + LG_{0,-1}$ | G |
| . | . | . | . | . | . | . | . | . |
| . | . | . | . | . | . | . | . | . |
| Z | $LG_{0,10}$ | 1 | 1 | 0 | 0 | 1 | $LG_{0,3} + LG_{0,-2} + LG_{0,1}$ | Z |
| 0 | $LG_{0,11}$ | 1 | 1 | 0 | 1 | 0 | $LG_{0,3} + LG_{0,-2} + LG_{0,-1}$ | 0 |
| 1 | $LG_{0,12}$ | 1 | 1 | 0 | 1 | 1 | $LG_{0,3} + LG_{0,-2} + LG_{0,-1} + LG_{0,1}$ | 1 |
| 2 | $LG_{0,13}$ | 1 | 1 | 1 | 0 | 0 | $LG_{0,3} + LG_{0,-2} + LG_{0,2}$ | 2 |
| 3 | $LG_{0,14}$ | 1 | 1 | 1 | 0 | 1 | $LG_{0,3} + LG_{0,-2} + LG_{0,2} + LG_{0,1}$ | 3 |
| 4 | $LG_{0,15}$ | 1 | 1 | 1 | 1 | 0 | $LG_{0,3} + LG_{0,-2} + LG_{0,2} + LG_{0,-1}$ | 4 |
| 5 | $LG_{0,16}$ | 1 | 1 | 1 | 1 | 1 | $LG_{0,3} + LG_{0,-2} + LG_{0,2} + LG_{0,1} + LG_{0,-1}$ | 5 |

Figure B. 5-bit information encoding schemes using (I) OAM Shift-Keying, (II) OAM Multiplexing

### 1.4 Astigmatic Transformed Far-field Speckle Patterns

Speckle-learned demultiplexing simplifies the need for precise beam shaping and detection of specific OAM modes by focusing on capturing the statistical properties of speckle patterns. This approach offers increased robustness, expedites computations, and remarkably, introduces alignment-free capabilities in free-space optical communication[14]. But we were limited to only one-half of the Orbital Angular Momentum (OAM) spectrum, which hindered us from utilizing the full spectrum. This limitation had a direct impact on the accuracy and spectral efficiency of our work, as higher-order OAM modes are more susceptible to noise and atmospheric perturbations.

The utilization of intensity-degenerate or mutually conjugate modes requires a means to break the degeneracy in order to recognize the modes through their distinctive speckle patterns. To achieve this, we introduce a tilted lens into the system. Lenses with circularly non-symmetric aberrations, such as the tilted spherical lens play a crucial role in disrupting the symmetry of the OAM field. This intentional disruption of symmetry results in the generation of far-field intensity patterns that exhibit unique characteristics. These distinctive patterns are directly tied to the specific OAM beam and its associated topological charge. As a result, the far-field intensity pattern produced can be effectively employed to identify the OAM beam and its corresponding charge[17-18].

Assuming the propagation direction of the optical wave is along the $z$-axis, the input field at the $z = 0$ plane, denoted as $U_i(x, y; 0)$, can be used to determine the corresponding propagated field at a parallel $z = d$ plane using the Fresnel diffraction integra given by

$$U(x, y; d) = F^{-1}[F(U_i(x, y; 0) \times H(f_x, f_y)] \tag{3}$$

where $H(f_x, f_y)] = exp[-i\pi\lambda d(f_x^2 + f_y^2)]$ is the free space Fresnel transfer function, $F$ and $F^{-1}$ are Fourier transform and inverse Fourier transform. $f_x$ and $f_y$ are the spatial frequencies whose values are given as $f_x = x/\lambda d$ and $f_y = y/\lambda d$ respectively.

To induce astigmatism into the far-field speckle pattern, we employed a tilted spherical lens at an angle β with respect to the $y$-axis. The phase transformation function of this tilted lens, which has a focal length of $f$, can be described as follows:

$$t_l(x, y) = exp(\frac{-ikx^2}{2f_1} + \frac{-iky^2}{2f_2}) \tag{4}$$

where $f_1$ and $f_2$ are the effective focal lengths in the $x$-axis and $y$-axis respectively whose values are given by $f_1 = f cos\beta$ and $f_2 = f / cos\beta$

The astigmatic speckle field in the far field region for intensity degenerate modes are calculated as

$$U_{sf}(x, y; f) = F^{-1}[F(U_i(x,y;0) \times H(f_x, f_y)] \tag{5}$$

where $U_i(x, y; 0) = U_{p.l}(\rho,\varphi,0) \times exp(i\varphi R) \times t_l(x, y)$

The random phases $\varphi_R$ are characterized by uniformly distributed phase values ranging from 0 to 2π. The intensity of the resulting astigmatic transformed far-field speckle pattern can be expressed as:

$$I_{sf}(x, y; f) = U_{sf}(x, y; f) \times U_{sf}(x, y; f)^* \tag{6}$$

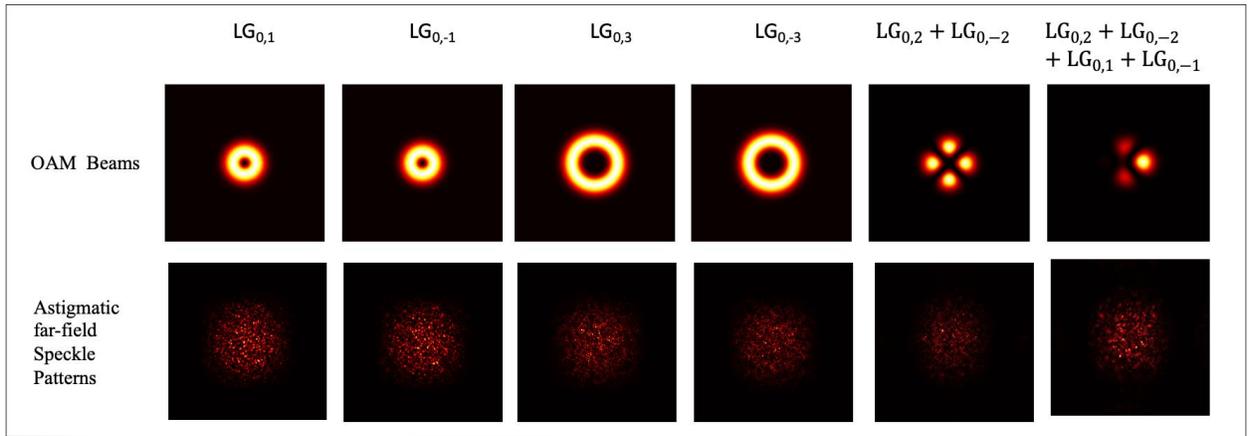

Figure C. OAM beams with their respective astigmatic far-field speckle patterns produced after passing through a tilted spherical lens for $\beta = 40°$.

**Speckle-learned Convolutional Neural Network**

For our study on OAM shift keying, we utilized 32 $LG_{p,l}$ modes including sixteen pairs of intensity-degenerate LG modes, ranging from $LG_{0,-16}$ to $LG_{0,16}$ (excluding $LG_{0,0}$). In OAM multiplexing, we used two pairs of intensity degenerate LG modes ($LG_{0,\pm2}$ & $LG_{0,\pm3}$) and $LG_{0,3}$ (excluding $LG_{0,0}$) to generate 32 superposed $LG_{p,l}$ modes. To train the CNN for modal demultiplexing, we simulated 1000 speckle pattern images of a resolution of $512 \times 512$ pixels for each mode.

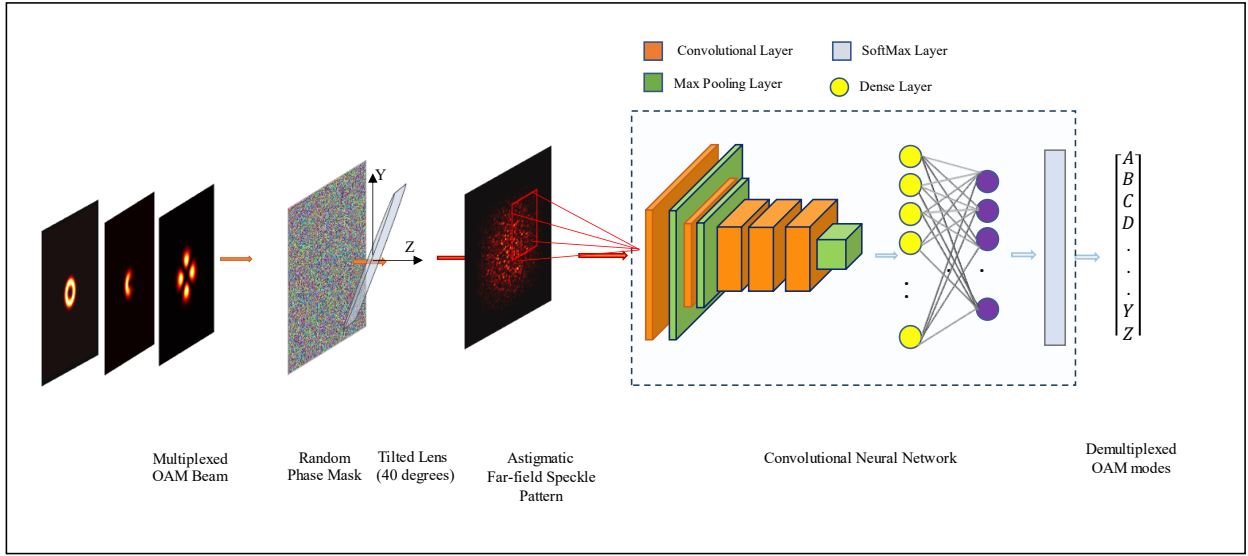

Figure D. The schematic diagram of the demultiplexing scheme using Speckle learned Convolutional Neural Network

The intensity speckle images ($I_{sf}$) were fed into an Alexnet[20], a pre-trained CNN known for its high accuracy and low computational load. This network includes five convolutional layers and three fully connected layers. We modified only the last layer of Alexnet to classify 32 modes, retaining its weights and biases, which were updated during training. The use of the pre-trained network reduced training time and computational load compared to training from scratch. We employed the "Stochastic Gradient Descent with Momentum" (SGDM) algorithm with a constant learning rate of 0.0001 for training. Out of the 1000 images in each class, 80% of the data were randomly selected for training, and the rest for model evaluation.

Now 32 intensity degenerate LG and intensity degenerate LG-SP modes are chosen for encoding alphabets from A to Z and digits 0 to 5. To demonstrate the proof of concept for its practical application, we have simulated an optical communication link where a phrase consisting of 'x' letters and 'y' numbers is encoded in superposed intensity degenerate Laguerre Gaussian modes. The propagated LG-SP modes are decoded using a speckle-based demultiplexing method, and the speckle-learned CNN has successfully reconstructed the encoded phrase with a classification accuracy of 100%.

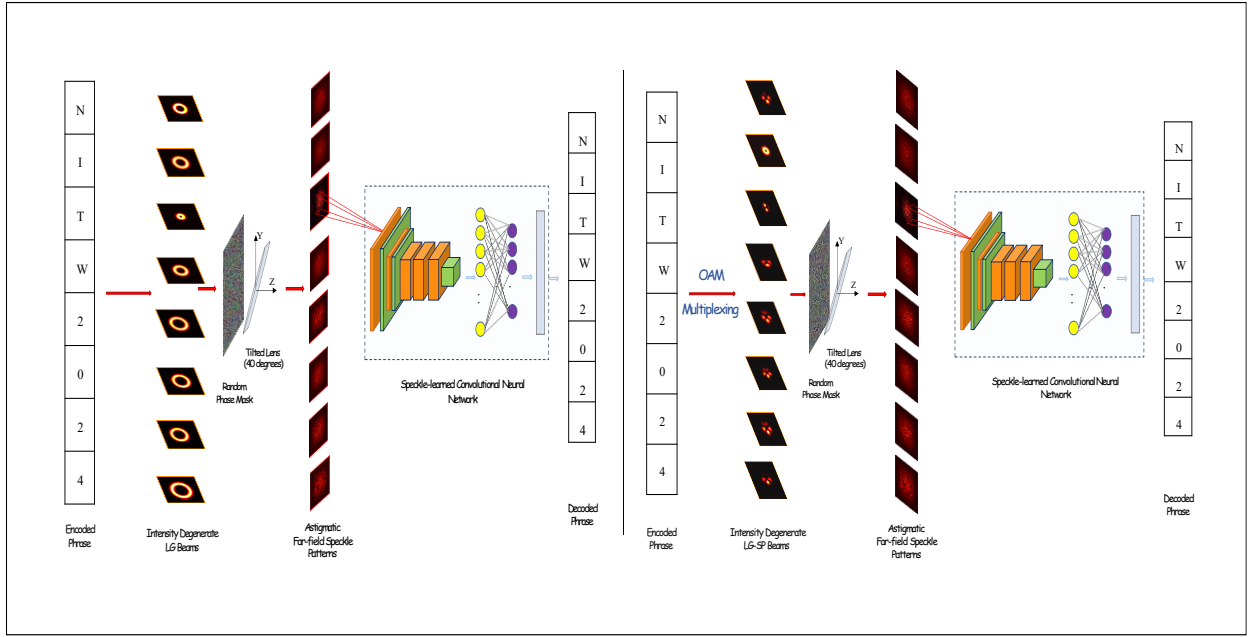

Figure E. Encoded and Decoding phrase "NITW2024" using OAM-SK and OAM Multiplexing.

## Result and Analysis

We demonstrated a speckle-learned convolutional neural network (CNN) for the recognition of intensity-degenerate orbital angular momentum (OAM) modes. These intensity-degenerate modes were then used to illustrate two distinct schemes, OAM shift keying and OAM multiplexing, in the context of real-world optical communication scenarios.

In the idealistic setting of our simulations, both OAM shift keying and OAM multiplexing exhibited impressive classification accuracy, exceeding 98%. However, it's important to acknowledge that in real-world applications, the situation becomes more complex. As the order of LG beams increases, the beam size also grows proportionally, with higher-order modes characterized by $LG_{p,l}$ having spot sizes that increase as $\sqrt{(l+2p+1)}$ [21]. These higher-order modes distribute their energy over larger rings with on-axis nulls, making them more susceptible to atmospheric turbulence and perturbations.

In the case of OAM shift keying, we employed eight pairs of intensity-degenerate LG modes, generating a total of 32 pure LG modes. In contrast, OAM multiplexing used five LG modes, to generate a set of 32 superposed intensity-degenerate $LG_{p,l}$ modes. Thus, it becomes evident that OAM multiplexing presents a more advantageous approach for real-world applications. This is primarily due to its use of lower-order intensity-degenerate LG modes to encode information. By incorporating lower-order intensity-degenerate modes, OAM multiplexing offers enhanced resilience against atmospheric turbulence and perturbations, which can significantly affect the propagation of higher-order modes. Given this, OAM multiplexing emerges as a more practical and preferable choice when compared to OAM shift keying in real-world scenarios.

## Summary

Our research explores the potential of orbital angular momentum Orbital angular momentum (OAM) beams in free-space optical communication. We introduce a novel approach to harness the entire spectrum of OAM modes, addressing noise and alignment challenges. By utilizing a speckle-learned deep learning model and introducing astigmatism in far-field speckle patterns through a tilted spherical convex lens, we achieve accurate differentiation of intensity-degenerate OAM modes. We further analyze Demultiplexing and Shift Keying techniques, showcasing their capability to identify mutual conjugate OAM modes. Both methods demonstrate >98% accuracy, making them effective for data transmission. In practical situations, the susceptibility of higher-order LG beam modes to atmospheric turbulence favors OAM Multiplexing as the better choice as a communication method. This approach ingeniously amalgamates the utilization of lower-order modes with the integration of intensity degenerate modes, resulting in a substantial enhancement of channel capacity, information transfer rates, and overall communication system reliability.

Overall, this paper provides a significant contribution to the field of speckle-based information processing aided with artificial intelligence (AI)-based methodologies, highlighting the potential of deep learning techniques for advancing the performance of demultiplexing methods. The proposed approach can pave the way for the development of more efficient and accurate speckle-based information processing systems.

**Funding:** SERB India (SRG/2021/001375)